\documentclass[12pt]{article}
\usepackage{graphicx}

\begin{document}

 \title{Photon Distribution Function for
Long-Distance Propagation of Partially Coherent Beams through the
Turbulent Atmosphere}
\author{G.P. Berman $^{1}$\footnote{Corresponding
author: gpb@lanl.gov} and A.A. Chumak$^{1,2}$
\\[3mm]$^1$
 Los Alamos National Laboratory, Theoretical Division,\\ Los Alamos, NM 87545
\\[5mm] $^2$  Institute of Physics of
the National Academy of Sciences\\ pr. Nauki 46, Kiev-28, MSP
03028 Ukraine
\\[3mm]\rule{0cm}{1pt}}
\maketitle \markright{right_head}{LAUR-06-1605}

\begin{abstract}

The photon density operator function is used to calculate light beam propagation through turbulent
atmosphere. A kinetic equation for the photon distribution function is derived and solved using the
method of characteristics. Optical wave correlations are described in terms of photon trajectories that
 depend on fluctuations of the refractive index. It is shown that both linear and quadratic
disturbances produce sizable effects for long-distance propagation. The quadratic terms are shown
to suppress the correlation of waves with different wave vectors.

We examine the intensity fluctuations of partially coherent beams
(beams whose initial spatial coherence is partially destroyed). Our
calculations show that it is possible to significantly reduce the
intensity fluctuations by using a partially coherent beam. The
physical mechanism responsible for this pronounced reduction is
similar to that of the Hanbury-Braun, Twiss effect.

\end{abstract}







\section{Introduction}

The study of the interaction of light beams with a random media is
of great importance for applications in such areas as astronomy,
laser communication, laser radar systems, etc. Fluctuations of the
atmospheric refractive index, caused by turbulent eddies, affect
electromagnetic waves. The effects of turbulence have been
investigated both theoretically and experimentally over the past few
decades. (See, for example, the monographs \cite{tatar}-\cite{andr}
and the surveys \cite{fant75}-\cite{kravt}.) Additional beam
spreading, beam wandering (``dancing"), intensity fluctuations
(known as ``scintillations"), etc. caused by turbulence limit
significantly the range and the performance of free-space
communication systems.

It has been established that coherent laser beams are very sensitive
to atmospheric conditions. Realizing this, the idea of utilizing of
partially coherent beams (PCB) for practical purposes has arisen. It
was shown by numerous researchers (see, for example, publications
\cite{fant79}-\cite{salem}) that partially coherent beams are less
affected by turbulence than fully coherent beams. A specific case
(in which the spatial coherence of the signal-carrying laser beam is
partially destroyed before it is launched into the atmospheric
channel) was considered by many researchers. This beam has the
angular spread in free space of the order of $\lambda /l_c$
($\lambda $ and $l_c$ are the wavelength and the transverse
correlation length of phase distortion at the source aperture,
respectively), which is larger than that of the coherent beam. For
not too large distances of propagation, beam spreading due to the
small initial coherence length, $l_c$, may dominate throughout the
trajectory. One can say that the effects of turbulence are masked by
the larger initial free-space spreading. At the same time it is
evident that, with increasing propagation distance, the atmospheric
inhomogeneity becomes the dominating factor. This results in a beam
size that is almost independent of the initial correlation length.

The dependence of the intensity fluctuations on the initial
coherence is more intricate than the dependence of beam spreading.
Provided the dependence of the intensity fluctuations on the initial
coherence is given, the noise/signal ratio can be controlled by
choosing the optimal initial coherence length, $l_c$. This tempting
opportunity has stimulated many studies in this field \cite{pole98},
\cite{jenn}-\cite{korot04}. It was shown that control is possible
for small fluctuations (weak turbulence or short-distance
propagation).

At  moderate and strong fluctuations the situation is more complicated.
Banakh {\it et al.} \cite{bana54} have shown that for long distances
of  propagation or for strong turbulence the intensity fluctuations
have a tendency to saturate at a level which  depends on the initial
spatial coherence. This level may be much lower than that of the
coherent beam by the factor, $(r_0/l_c)^2$, where $r_0$ is the
initial radius of the beam. The  paper deals with the physical model
when the measuring device (detector) has the response time, $\tau
_d$, greater than the coherence time, $\tau _s$, of the radiation at
the beginning of the trajectory. In this case, the detector averages
the signal over the initial fluctuations, which are due to temporal
intensity fluctuations of the source and (or) may be generated when
a coherent beam is transformed into PCB.  The authors of
\cite{bana54} have solved an equation for the fourth-order coherence
function (the fourth moment of the field). The derivation of the
equation for fourth moment in the limit of Markov approximation is
given in \cite{tatar}, \cite{fant75}. Its solution is based on the
approximate method developed by Yakushkin in \cite{yaku75}. The
authors of \cite{bana54} have modified the Yakushkin approach to be
applicable to the case of PCB. The Yakushkin method is based on the
observation that, for long distance propagation, the dominant
contribution to the fourth-order correlation function comes from the
products of two second-order coherence functions just as if Gaussian
statistics for radiation field were valid. The Dashen analysis
\cite{dash} in terms of the path integral formalism and the results
of Fante \cite{Rfant80} obtained by employing the extended
Huygens-Fresnel principle have justified the Yakushkin idea. It
seems to be quite reasonable to consider that after long-distance
propagation through a medium with a random refractive index, the
statistics of the photon flux acquires some of the properties of
thermal radiation.

For similar conditions, when the coherence time of the
quasimonochromatic laser source is much smaller than the detector's
integration time interval, $\tau_d$,  Korotkova {\it et al.} \cite{korot04}
have derived an analytical expression (see Eq. (33) in \cite{korot04})
for the  intensity fluctuations. The paper \cite{korot04} has
generalized the analytical approach developed by Andrews and
Phillips (see monographs \cite{anphi} and \cite{andr}) to the case
of PCB. This approach is a modification of the Rytov theory. Namely,
in addition to the first-order perturbation of the complex phase of
the field caused by refractive index fluctuations (as in Rytov
theory), second-order perturbation terms are taken into account.
The results of \cite{korot04} differs from those in \cite{bana54}.
More comments are presented in Section 4.

The purpose of  our  paper is to develop a new \emph{quantum}
perturbation theory based on path integrals. This approach allows us
to obtain the scintillation index for the case of a PCB, including a
limit of a strong turbulence and long-distance propagation. Our
approach uses the distribution function of photons (for
photon density in phase space), $f({\bf r},{\bf q},t)$, where ${\bf
r}$ and ${\bf q}$ are the coordinate and the momentum of photon, to
describe the beam characteristics at an arbitrary instant, $t$. The
first and second moments of the distribution function  ($<f>$ and
$<ff>$) are used to obtain the beam size and intensity fluctuations,
respectively. Also, higher-order moments will be obtained for long
propagation paths.

The next Section deals with the definition of the distribution
function. The equation governing the evolution of the distribution
function will be derived.

\section{The photon distribution function and its evolution}

The Hamiltonian of photons in a medium with a fluctuating refractive index is given by
\begin{equation}\label{one}
H=\sum_{\bf k}\hbar \omega _{\bf k}b^+_{\bf k}b_{\bf k}-
\sum_{{\bf k},{\bf k^\prime}}\hbar \omega _{\bf k}n_{\bf k^\prime}b^+_{\bf k}b_{{\bf k}
+{\bf k^\prime}},
\end{equation}
where the two terms on the right-hand side describe photons in a vacuum and
the effect of refractive index fluctuations, respectively; $b^+_{\bf k}$ and
$b_{\bf k}$ are creation and annihilation operators of photons with momentum
${\bf k}$, $\hbar \omega _{\bf k}\equiv \hbar ck$ is the photon energy,
$c$ is the speed of light
in a vacuum, and $n_{\bf k}$ is the Fourier transform of the refractive index
fluctuations $\delta n({\bf r})$. The Fourier transform is defined by
\begin{equation}\label{two}
n_{\bf k}=\frac 1V\int dVe^{i{\bf kr}}\delta n({\bf r}),
\end{equation}
were $V\equiv L_xL_yL_z$ is the normalizing volume.

Eq. (\ref{one}) follows from the representation of the energy of the
random medium + electromagnetic field given in \cite{land} (Chapter
15) in the limit of small wave-vectors ${\bf {k^{\prime}}}$
($k^{\prime}\ll k$) and of atmosphere refractive index close to
unity ($n({\bf r})-1<<1$). The former means that the scale of
spatial inhomogeneity of turbulence is much greater than the
wavelength of the radiation. For simplicity, we consider here only
the case of polarized light with a fixed polarization throughout the
distance of propagation. Depolarization effects due to atmosphere
turbulence are very small. (See, for example, papers \cite{stroh}
and \cite{coll}.) Also, the terms describing the zero-point
electromagnetic energy are omitted in Eq. (\ref{one}).

The photon distribution function is defined by
analogy with the distribution functions for electrons, phonons \cite{tomc},
etc. It is given by
\begin{equation}\label{three}
f({\bf r},{\bf q},t)=\frac 1V\sum_{\bf k}e^{-i{\bf kr}}b^+_{{\bf q}+
{\bf k}/2}b_{{\bf q}-{\bf k}/2},
\end{equation}

The distribution function will be used to describe beams with characteristic sizes much
larger than the photon wave-length. Thus, we will restrict a sum
over ${\bf k}$ by some $k_0$ ($k<k_0<<q_0$, where $q_0$
is the wave vector corresponding to the central frequency $\omega _0$ of radiation,
$\omega _0=cq_0$).
At the same time $k_0$ is chosen to be large enough to sample the spatial variation of the
light intensity.

 After integration of the operator function $f({\bf r},{\bf q},t)$
over the volume $V$, we obtain the operator for the total number of photons
with momentum ${\bf q}$:
\begin{equation}\label{four}
\int dVf({\bf r},{\bf q},t)=b^+_{\bf q}b_{\bf q}.
\end{equation}
Similarly, the quantity obtained after a summation of $f({\bf
r},{\bf q},t)$ over ${\bf q}$ may serve as a photon density averaged
over a small spatial area with the size $\pi /k_0$. This is very
similar to the Mandel operator \cite{mand} (Chapter 12) introduced
in \cite{mand66}.

Here and in the remainder of this paper, we use the Heisenberg
representation for all operators. Thus, the evolution of $f({\bf
r},{\bf q},t)$ is determined by the commutator with the total
Hamiltonian:
\begin{equation}\label{five}
\partial_t f{\bf r},{\bf q},t)=\frac 1{i\hbar }[f{\bf r},{\bf q},t),H].
\end{equation}
Eq. (\ref{five}) can be written explicitly as
\begin{equation}\label{six}
\partial_t f({\bf r},{\bf q},t)+{\bf c_q}\partial_{\bf r}f({\bf r},{\bf q},t)-
i\omega _0\sum_{\bf k}e^{-i{\bf kr}}n_{\bf k}\Bigg[f\Bigg({\bf r},{\bf q}+\frac {\bf k}2,t\Bigg)-
f\Bigg({\bf r},{\bf q}-\frac {\bf k}2,t\Bigg)\Bigg]=0,
\end{equation}
where ${\bf c_q}=\partial_{\bf q}\omega _{\bf q}$.

Considering the characteristic values of the photon momentum to be much greater than the wave
vectors of turbulence, we can express the difference of functions in square brackets
by the corresponding derivative. A detailed discussion of this approximation is given in Section 6.
Then, after summing over ${\bf k}$, we obtain
\begin{equation}\label{seven}
\{ \partial_t +{\bf c_q}\partial_{\bf r}+{\bf F}({\bf r})\partial_{\bf q}\}
f({\bf r},{\bf q},t)=0,
\end{equation}
where ${\bf F}({\bf r})=
\omega _0\partial_{\bf r}n({\bf r})$.
As we see, the photon distribution function is governed by a kinetic equation in which
the random force ${\bf F}({\bf r})$ originates from atmospheric turbulence.

The distribution function determines density of photons at a point (${\bf r}$,${\bf q}$)
of the phase space at time $t$. Eq. (\ref{seven}) may be interpreted as the equation
governing the evolution of a particle distribution function in which the state of each particle is
described by the individual coordinate ${\bf r}$ and the momentum ${\bf q}$. The
trajectories of these particles may be obtained from the solution of the equations of motion:
\[ \frac {\partial {\bf r}(t)}{\partial t}={\bf c}({\bf q}(t)), \]
\begin{equation}\label{eight}
\frac {\partial {\bf q}(t)}{\partial t}={\bf F}({\bf r}(t)).
\end{equation}
Then, the general solution of Eq. (\ref{seven}) is given by
\begin{equation}\label{nine}
f({\bf r},{\bf q},t)=\phi \Bigg\{{\bf r}-\int _0^tdt^\prime\frac {\partial
{\bf r}(t^\prime)}{\partial t^\prime};{\bf q}-\int_0^tdt^\prime\frac {\partial
{\bf q}(t^{\prime})}{\partial t^\prime}\Bigg\},
\end{equation}
where the function $\phi ({\bf r},{\bf q})$ is the ``initial" value
of $f({\bf r},{\bf q},t)$, i.e.
\begin{equation}\label{ten}
\phi ({\bf r},{\bf q})=\frac 1V\sum_{\bf k}e^{-i{\bf kr}}(b^+_{{\bf q}+
{\bf k}/2}b_{{\bf q}-{\bf k}/2})|_{t=0}\equiv \sum_{\bf k}e^{-i{\bf kr}}
\phi ({\bf k},{\bf q}),
\end{equation}
and the ``trajectories" ${\bf r}(t^\prime )$ and ${\bf q}(t^\prime)$
pass through the point ${\bf r},{\bf q}$ at $t^\prime =t$ [i.e.
${\bf r}(t^\prime =t)={\bf r}, {\bf q}(t^\prime =t)={\bf q})$]. As
one can see, the photon distribution function at an arbitrary
instant $t$ is expressed via the operators $b^+_{\bf q},b_{\bf q}$
defined for some fixed $t_0$ ($t_0$ is chosen to be equal to $0$ in
Eq. (\ref{nine}). It is convenient to put $t-t_0=z/c$. Thus, $t_0$
is the instant when photons exit from the source. The initial photon
statistics (at $t_0$), determined by the source properties, is
assumed to be given.

We consider here the propagation of light beams with narrow spread
(paraxial beams). In this case, $k_{\bot } , q_{\bot }\ll q_0$,
where index ($_\bot $) means perpendicular to the direction of
propagation (the $z$-axis) components. The relative effect of
turbulence on $q_z$ is negligible because of the large value of
$q_0$. At the same time, $q_\bot $, which determines a beam
divergence, can be increased considerably due to turbulence
(compared to the initial value). Therefore, beam characteristics
should be modified significantly for the case of long distance
propagation.

It follows from Eq. (\ref{eight}) that the evolution of transverse photon momentum is
given by
\begin{equation}\label{eleven}
{\bf q}_\bot (t^\prime )={\bf q}_\bot +\int _t^{t^{\prime }}dt^{\prime \prime }
{\bf F}_{\bot }[{\bf r}(t^{\prime \prime})].
\end{equation}
Similarly, we obtain an expression for ${\bf r}(t^\prime )$
\begin{equation}\label{twelve}
{\bf r} (t^\prime )={\bf r}-{\bf c _q }(t-t^\prime) -\frac c{q_0}
\int _t^{t^{\prime }}dt^{\prime \prime }(t^{\prime \prime}-t^{\prime })
{\bf F}_{\bot }[{\bf r}(t^{\prime \prime})].
\end{equation}
Then, Eq. (\ref{nine}) can be written as
 \begin{equation}\label{thirteen}
f({\bf r},{\bf q},t)=\phi \Bigg\{{\bf r}-{\bf c_q}t+\frac c{q_0}\int _0^tdt^\prime
t^\prime {\bf F}_\bot [{\bf r}(t^\prime )];{\bf q}-\int _0^tdt^\prime
{\bf F}_\bot [{\bf r}(t^\prime )]\Bigg\}.
\end{equation}

A regular iterative procedure is applicable here to expand ${\bf r} (t^\prime )$
 in powers of ${\bf F}$. Then, substituting the explicit
terms ${\bf r} (t^\prime )$ into Eq. (\ref{thirteen}), we will
obtain solution of the problem. In particular, the first and second
moments of $f$, which describe beam spreading and intensity
fluctuation, can be calculated. Applying this perturbation method,
we can investigate the effects of the initial partial (spatial)
coherence on beam spreading and scintillations.

\section{Beam spread and intensity fluctuations}

The intensity of radiation in the $z$-direction at ${\bf r}$ can be presented in the form
\begin{equation}\label{fourteen}
I({\bf r})=c\sum_{\bf q}\hbar \omega _{\bf q}f({\bf r},{\bf q},t).
\end{equation}
This can be rewritten as
\begin{equation}\label{fifteen}
I({\bf r})=c\hbar \omega _0\sum_{\bf q,k}e^{-i{\bf k}\{ {\bf r}-{\bf c}({\bf q})t+
(c/q_ 0)\int _0^tdt^\prime t^\prime {\bf F}_{\bot }[{\bf r}(t^{\prime})]\} }
\phi _{\bf k}\Bigg\{ {\bf q}-\int _0^tdt^\prime t^\prime {\bf F}_\bot
[{\bf r}(t^{\prime })]\Bigg\}.
\end{equation}
Let us restrict ouselves to only linear in ${\bf F}$ terms in Eq. (\ref{twelve}).
In other words, we put ${\bf r}-{\bf c _q }(t-t^\prime)$ for
${\bf r}(t^{\prime })$ in arguments of ${\bf F}_\bot $ of Eq. (\ref{fifteen}). After
changing variables ${\bf q}_\bot -\int _0^tdt^\prime t^\prime {\bf F}_\bot
[{\bf r}(t^{\prime })]\rightarrow {\bf q}_\bot $ and using the relation
$z=ct$, Eq. (\ref{fifteen}) is transformed to
\begin{equation}\label{sixteen}
I({\bf r})=c\hbar \omega _0\sum_{\bf q,k}e^{-i{\bf k}_\bot \{ {\bf r}_\bot -
{\bf q}_\bot (z/q_0)+
(c/q_0)\int _0^tdt^\prime (t-t^\prime ){\bf F}_{\bot }[{\bf r}-c({\bf q})(t-t^
\prime )]\} }\phi _{\bf k}({\bf q}).
\end{equation}
The stochastic variables ${\bf F}$ and $\phi _{\bf k}({\bf q})$,
which are of different nature, are separated in Eq. (\ref{sixteen}).
Averaging of each factor in the sum can be performed independently
because of the absence of correlations between the source
fluctuations and the refractive index fluctuations. Thus, we have
\begin{equation}\label{seventeen}
<I({\bf r})>=c\hbar \omega _0\sum_{\bf q,k}<e^{-i{\bf k}_\bot \{ {\bf r}_\bot -
{\bf q}_\bot (z/q_0)+
(c/q_0)\int _0^tdt^\prime t^\prime {\bf F}_{\bot }[{\bf r}-c({\bf q})t^\prime ]
\} }><\phi _{\bf k}({\bf q})>.
\end{equation}
In Eq. (\ref{seventeen}) and throughout this paper, we shall calculate average
values of functionals of $\delta n$. This can be carried out when the statistics of
$\delta n$
is known. Usually, $\delta n$ is assumed to be a Gaussian random variable with known
covariance $<\delta n({\bf r})\delta n({\bf r}^\prime )>$. The covariance is defined
by its Fourier transform, $\psi ({\bf g})$, with respect
to the difference ${\bf r}-{\bf r}^\prime$. The dependence $\psi ({\bf g})$ is often
approximated by the von Karman formula
\begin{equation}\label{eighteen}
\psi ({\bf g})=0.033C_n^2\frac {exp[-(gl_0/2\pi
)^2]}{[g^2+L_0^{-2}]^{11/6}}.
\end{equation}
The quantities $L_0$ and $l_0$ are the outer and inner scales sizes
of the turbulent eddies, respectively. In atmospheric turbulence,
$L_0$ may range from 1 to 100 meters, and $l_0$ is usually on the
order of several millimeters. $C_n^2$ is known as the
index-of-refraction structure constant. In most physically important
cases the quantity $L_0^{-2}$ in the denominator of Eq.
(\ref{eighteen}) can be omitted. In this case, the von Karman
spectrum is reduced to the Tatarskii spectrum \cite{tatar}.

Using the explicit form for the turbulence fluctuations, Eq.
(\ref{seventeen}) becomes
\begin{equation}\label{nineteen}
<I({\bf r})>=c\hbar \omega _0\sum_{\bf q,k}e^{-i{\bf k}_\bot [{\bf r}_\bot -
{\bf q}_\bot (z/{q_0})]-k^2_\bot z^3T}<\phi _{\bf k}({\bf q})> ,
\end{equation}
where the effect of turbulence is represented by a quantity $T$
 ($T=0.558C_n^2l_0^{-1/3}$). When obtaining Eq. (\ref{nineteen}) we
have assumed that the distance of propagation is much greater than
the characteristic length of the turbulence. The condition $z>>L_0$
is sufficient to satisfy this requirement for any regime of
propagation. Also, the Tatarskii expression for the turbulence
spectrum was used.

Now we consider the average value of $<\phi _{\bf k}({\bf q})>$. It depends
entirely on the source properties.
Let us consider a source field with the following mode structure
\begin{equation}\label{twenty}
E^{s}({\bf r})=\sum_n\Bigg(\frac {2\pi \hbar \omega _n}{L_z}\Bigg)^{1/2} [ e^{iq_nz}\Phi_
n({\bf r}_\bot )b_n+e^{-iq_nz}\Phi ^*_n({\bf r}_\bot )b^+_n],
\end{equation}
where the normalized function, $\Phi ({\bf r}_\bot )$, wave vector, $q_n$, and frequency,
$\omega _n$, describe
the profile and eigenfrequency of $n$-th mode, respectively. This field should be
matched with the field in the atmosphere,
\begin{equation}\label{two}
E^{atm}({\bf r})=\sum_{\bf q}\Bigg(\frac {2\pi \hbar \omega _{\bf
q}}{V}\Bigg)^{1/2} [e^{i{\bf qr}}b_{\bf q}+e^{-i{\bf qr}}b^+_{\bf
q}],
\end{equation}
in the plane of the transmitter, where functions $\Phi _n({\bf r}_\bot )$ are assumed
to be known. When we deal with single-mode laser radiation (for example, with the $n=0$
 mode), only one term in the sum in Eq. (\ref{twenty}) should be retained. The other terms
can be omitted. It is important to point out that rigorous matching
conditions would also involve the vacuum fields to maintain correct commutation relations for
all field operators. Nevertheless, the vacuum fields may be neglected in our case for
two reasons:
(i) the photon detectors are assumed to be of the absorbing type, hence they are not
sensitive to vacuum fields, and (ii) we consider only linear (in $E$) propagation
of the radiation.
In this case,
\begin{equation}\label{twt}
b_{{\bf q}_\bot ,q_0}=b(L_xL_y)^{-1/2}\int d{\bf r}_\bot e^{-i{\bf q}_\bot
{\bf r}_\bot }\Phi ({\bf r}_\bot ),
\end{equation}
where the index in $\Phi $ is dropped for brevity.

Until now, all phase distortions were not considered. In practice, stochastic phase
distortions may be introduced by means of a rotating phase diffuser placed in front of the
aperture.
 Mathematically, the effect of the phase diffuser may be taken into
account by introducing the multiplier $e^{-i\varphi ({\bf r}_\bot )}$ (see, for example,
\cite{bayk85})
into the integrand of Eq. (\ref{twt}), where $\varphi ({\bf r}_\bot )={\bf a}{\bf r}_\bot $
and ${\bf a}$ is a Gaussian random variable with covariance
$<(a_{x,y})^2>=\lambda _c^{-2}$. Then, considering $\Phi $ to be a Gaussian-type function
($\Phi =\frac {\sqrt 2}{{\sqrt \pi }r_0}e^{-r^2_\bot /r_0^2}$), we obtain
\begin{equation}\label{twth}
<\phi _{\bf k}({\bf q})>=\frac {2\pi r_1^2}{VL_xL_y} <b^+b>e^{-k^2(r_0^2/8)-q^2(r_1^2/2)},
\end{equation}
where $r^2_1=r_0^2/(1+2r_0^2\lambda _c^{-2})$ and $<b^+b>=|\beta|^2$ for the coherent state
$|\beta >$ of the laser radiation. Here, symbols ${\bf q}$ and ${\bf k}$ are the perpendicular
components of the wave vectors.
As we see, the effect of partial coherence is represented by the value of $r_1^2$. In
the limiting case of $\lambda _c\rightarrow \infty $, $r_1^2$ tends to  $r_0^2$. In the
opposite case of small correlation length $\lambda _c\rightarrow 0$,
$r_1^2$ tends to $\lambda ^2_c/2$. Hence, the quantity $b=r_1^2/r_0^2$ is a measure of a
spatial coherence of the beam.

Using Eqs. (\ref{nineteen}) and (\ref{twth}), it is found that
\begin{equation}\label{twf}
<I(z,{\bf r}_\bot =0)>=I_0\Bigg[1+\frac {4z^2}{q_0^2r_0^2r_1^2}+\frac {8z^3T}{r_0^2}\Bigg]^{-1},
\end{equation}
where $I_0$ is equal to $<I({\bf r})>$ at ${\bf r}_\bot =0$ and $z=0$. Eq. (\ref{twf})
coincides
with the corresponding Eq. (39) of \cite{fant75} when $C_n^2$ does not depend on distance $z$
and $r_1=r_0$.
\begin{figure}[ht]
\centering
\includegraphics{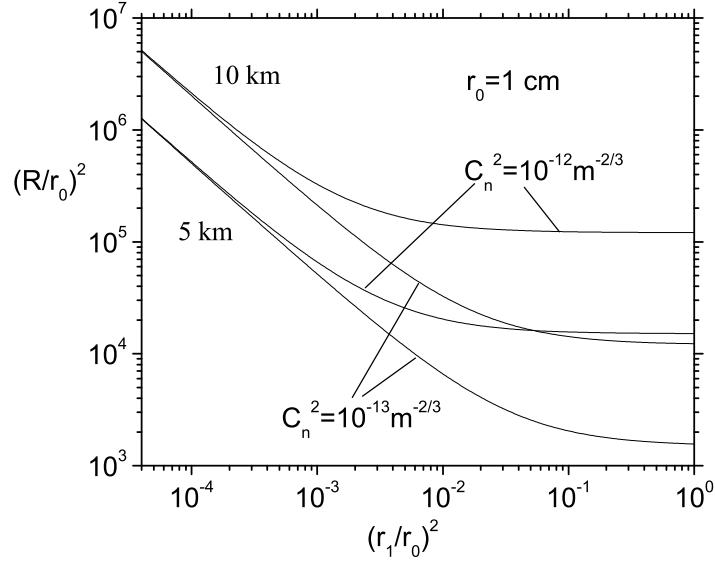}
\caption{The dependence of beam radius on the coherence length parameter $(r_1/r_0)^2$ for
$q_0=10^7m^{-1},(l_0/2\pi)=10^{-3}m$. There is a
wide range of $(r_1/r_0)^2$ for which the beam radius is almost constant.}
\label{one}
\end{figure}

The intensity for arbitrary $\bf r  _\bot$ can be obtained from (Eq.
\ref{twf}) by multiplying its right-hand side by the factor
$exp\{-\frac {2r_\bot ^2}{r_0^2}[1+\frac {4z^2}{q_0^2r_0^2r_1^2}+
\frac {8z^3T}{r_0^2}]^{-1}\}$. The average beam radius $R $ defined
as,
\begin{equation}\label{twfi}
R^2=\frac {\int d{\bf r}_\bot r_\bot ^2<I(r_\bot )>}{\int d{\bf r}_\bot <I(r_\bot )>},
\end{equation}
is given by
\begin{equation}\label{tws}
R^2=\frac {r_0^2}2\Bigg[1+\frac {4z^2}{q_0^2r_0^2r_1^2}+\frac {8z^3T}{r_0^2}\Bigg].
\end{equation}
This coincides with Eq. (4) of the paper \cite{salem}, where the
effect of partial coherence was studied. As one can see, only the
second term in the square brackets depends on the initial coherence
via $r _1^2$. This term describes the diffraction spreading of the
beam in free space. It depends on both the initial beam radius
$r_0/{\sqrt 2}$ and the coherence length $l_c$ via $r_1$. The
spreading may be enhanced considerably, if $\lambda_c$ is
decreasing. In this way, the diffraction divergence may exceed the
divergence due to turbulence (the third term) for a broad range of
distances. In this case, we note the independence of the beam radius
on the turbulence strength, i.e. on the weather conditions.
Nevertheless, it follows from Eq. (\ref{twf}) that the ``turbulent"
term dominates in the limit of $z\rightarrow \infty$. Fig. 1 shows
the dependence $R^2(r_1^2/r_0^2)$ for two different distances and
turbulence strengths.

Prior to considering the intensity fluctuations, it is useful to
analyse qualitatively peculiarities of the wave correlations in the
course of their propagation through the atmosphere. It follows from
Eqs. (\ref{twth}) and (\ref{nineteen}) that the characteristic
values of $k$ contributing to $<I>$ are less than or of the order of
the smallest of the quantities $\frac {2{\sqrt 2}}{r_0}$, $\frac
{q_0r_1}{z{\sqrt 2}},(z^3T)^{-1/2}$. This means that the two waves,
$b _{{\bf q}+{\bf k}/2}^+$ and $b _{{\bf q}-{\bf k}/2}$, are
correlated ($<b _{{\bf q}+{\bf k}/2}^+b _{{\bf q}-{\bf k}/2}> {\not
=}0$) if $k$ satisfies the above requirement. In the limit of
$z\rightarrow \infty $, each wave correlates with itself only as if
the beam originates from a thermal source. The characteristic
distance for wave randomization can be obtained from the requirement
that the ``turbulent" term should be dominant term in Eq.
(\ref{tws}). The corresponding criteria are given by
\[8z^3T>r_0^2, 4z^2q_0^{-2}r_1^{-2}.\]
These inequalities distinguish the range of strong turbulence, which
is of the most interest for our studies.

The intensity fluctuations are determined by the expression
 \begin{equation}\label{twse}
<:I^2({\bf r}):>=\frac {(c\hbar \omega _0)^2}{V^2}\sum_{\bf q,k}\sum_{{\bf q}^\prime ,{\bf k}^\prime }
e^{-i({\bf k}+{\bf k}^\prime ){\bf r}}<b _{{\bf q}+{\bf k}/2}^+b _{{\bf q}^\prime+{\bf k}^\prime /2}^+
b _{{\bf q}^\prime -{\bf k}^\prime /2}b _{{\bf q}-{\bf k}/2}>,
\end{equation}
where the symbol $\{ :  :\}$ means the normal ordering of the
creation and annihilation operators. (See more detail in
\cite{mand}.)

Let us introduce the notation:
\[ G({\bf q},{\bf q}^\prime ;{\bf k},{\bf k}^\prime )\equiv
b _{{\bf q}+{\bf k}/2}^+b _{{\bf q}^\prime+{\bf k}^\prime /2}^+
b _{{\bf q}^\prime -{\bf k}^\prime /2}b _{{\bf q}-{\bf k}/2}.\]
In the limit $z\rightarrow \infty $ we have
\begin{equation}\label{twei}
<G>=<b _{{\bf q}+{\bf k}/2}^+b _{{\bf q}-{\bf k}/2}><b _{{\bf q}^\prime +{\bf k}^\prime /2}^+
b _{{\bf q}^\prime -{\bf k}^\prime /2}>
\end{equation}
\[ +<b _{{\bf q}+{\bf k}/2}^+b _{{\bf q}^\prime -{\bf k}^\prime /2}><b _{{\bf q}^\prime+{\bf k}^\prime /2}^+
b _{{\bf q}-{\bf k}/2}>=n_{\bf q}n_{{\bf q}^\prime}\delta _{{\bf
k},0}\delta _{{\bf k}^{\prime },0} +n_{{\bf q}+{\bf k}/2}n_{{\bf
q}-{\bf k}/2}\delta _{{\bf q},{\bf q}^\prime }\delta _{{\bf k},-
{\bf k}^\prime },\] where $n_{\bf q}\equiv <b^+_{\bf q}b_{\bf q}>$.
The terms which describe two pairs of waves with coinciding indices
in each pair, have nonzero values. At large but finite $z$,
correlation of waves with somewhat different indices (``nondiagonal"
terms) may also occur. These are terms with: (i) $k,k^\prime \leq
(z^3T)^{-1/2}$ or (ii) $|{\bf q}-{\bf q}^\prime +({\bf k}+{\bf
k}^\prime )/2|,|{\bf q}^\prime -{\bf q} + ({\bf k}+{\bf k}^\prime
)/2|\leq (z^3T)^{-1/2}$. The estimates, (i) and (ii), follow from
the requirement that the deviations of the wave-vectors from their
``diagonal" values are less than or of the order of the reduced beam
radius. The last condition is determined by the ``turbulent" term in
Eq. (\ref{tws}) for the case of long distance propagation. The
intersection of both ranges of wave-vectors confined by inequalities
(i) and (ii) may be neglected because of small volume in wave-vector
space. Then, Eq. (\ref{twei}) is reduced to
\begin{equation}\label{twn}
<:I^2({\bf r}):>=\frac {(c\hbar \omega _0)^2}{V^2}(\Sigma  _1+\Sigma  _2)<G({\bf q},{\bf q}^\prime ;
{\bf k},{\bf k}^\prime )>,
\end{equation}
where $\Sigma_{1,2}$ means summation, with the restrictions (i) and (ii), respectively.
The second sum is reduced to the first one by renaming the indices. Then considering the quantity
$(1/V)b^+_{{\bf q}+{\bf k}/2}b_{{\bf q}-{\bf k}/2}$ as a spatial Fourier component of the distribution
function, we have
\begin{equation}\label{thir}
\frac {<:I^2({\bf r}):>}{2(c\hbar \omega _0)^2}=\Sigma <e^{-i\{{\bf k}[{\bf r}-{\bf c}({\bf q})t]+
{\bf k}^\prime [{\bf r}-{\bf c}({\bf q}^\prime )t]+(c/q_0)\int _0^tdt^\prime
t^\prime ({\bf k}{\bf F}[{\bf r}({\bf q},t^\prime )]+{\bf k}^\prime {\bf F}[{\bf r}({\bf q}^\prime ,
t^\prime )])\}}
\end{equation}
\[ \times :\phi _{\bf k}\{{\bf q}-\int _0^tdt^\prime {\bf F}[{\bf r}({\bf q},t^\prime )]\}
\phi _{{\bf k}^\prime }\{{\bf q}^\prime -\int _0^tdt^\prime {\bf F}[{\bf r}
({\bf q}^\prime,t^\prime )]\}: >, \]
where the summation is with restriction (i).
  For the sake of brevity, we again denote by
${\bf F}$ the component of the force perpendicular to the $z$-axis.

The averaging of the product $:\phi \phi :$ with respect to source
variables can be performed straightforwardly. (See the derivation of
Eq. (\ref{twth}).) It is given by
\begin{equation}\label{thiro}
<:\phi _{\bf k}({\bf Q})\phi _{{\bf k}^\prime }({\bf Q}^\prime):>=
\Bigg(\frac {2\pi r_0r_2}{VL_xL_y}\Bigg)^2<b^+b^+bb>
\end{equation}
\[ \times e^{-({\bf Q}-{\bf Q}^\prime )^2r_0^2/4-({\bf Q}+{\bf Q}^\prime )^2r_2^2/4-(k^2+k^{\prime 2})r_0^2/8},\]
where $<b^+b^+bb>=|\beta |^4$,
\[{\bf Q}={\bf q}-\int _0^tdt^\prime {\bf F}[{\bf r}({\bf q},t^\prime )],
{\bf Q}^\prime ={\bf q}^\prime -\int _0^tdt^\prime {\bf F}[{\bf r}({\bf q}^\prime,t^\prime ),\]
 and the quantity $r_2$ is defined (similarly to $r_1$) as $r_2^2=r_0^2/(1+4r_0^2\lambda _c^{-2})$.

The rest of the calculations required for obtaining $<:I^2:>$ can be carried out according to the scheme
outlined for the case of $<I>$.

Up to now, we have dealt with the equal-time correlation function
$<:~I(t)I(t):>$. This analysis is relevant to the experimental
situation in which the response time of the detector, $\tau _d$, is
much less than the characteristic time of the phase diffuser, $\tau
_s$. In what follows, we consider the opposite case, $\tau _d>>\tau
_s$. The detector averages intensity fluctuations during the time
interval $\tau _d$. Lowering of the noise level may be expected for
this detector. Although the time interval, $\tau _d$, is much larger
than, $\tau _s$, at the same time it should be much shorter than the
characteristic time of the turbulence evolution (frozen turbulence),
i.e. $\tau _d<<\tau _a=l/v_a$, where $l$ and $v _a$ are the
characteristic radius of the turbulent eddies and their transverse
flow velocity across the beam, respectively. The average value
$<:I(t)I(t+\tau ):>$ with $\tau \sim \tau_d$ gives the dominant
contribution to the quantity measured by this detector. In this
case, the atmospheric conditions of light propagations may be
considered to be fixed, while the initial correlations of four field
operators $b_{\bf q}^+,b_{\bf q}$, which enter $<:I(t)I(t+\tau ):>$,
should be calculated accounting for the very different random phases
$\varphi (t_0)$ and $\varphi (t_0+\tau )$ introduced by the
diffuser. These phases do not correlate ($<\varphi ({\bf
r},t)\varphi ({\bf r}^\prime ,t+\tau)>=0$) even at ${\bf r}={\bf
r}^\prime $.  Hence, averaging over the random phases of each of the
product of four operators is reduced to calculations of $<e^{-i\{
\varphi [{\bf r}_1(t_0)]+\varphi [{\bf r}_2(t_0+\tau )]-\varphi
[{\bf r}_3(t_0+\tau )]- \varphi [{\bf r}_4(t_0)]\}}>$, that is equal
to $<e^{-i\{ \varphi ({\bf r}_1)-\varphi ({\bf r}_4)\} }> <e^{-i\{
\varphi ({\bf r}_2)-\varphi ({\bf r}_3)\} }>$. As a result, the
general expression for the intensity correlations will become
different from Eq. (\ref{thir}). The difference is that Eq.
(\ref{thiro}) should become
 \begin{equation}\label{thirt}
2\Bigg(\frac {\pi r_1^2}{VL_xL_y}\Bigg)^2<b^+b^+bb>\Bigg\{ e^{-(Q^2+{Q^\prime }^2)r_1^2/2-
(k^2+{k^\prime }^2)r_0^2/8}
\end{equation}
\[+e^{-[({\bf Q}-{\bf Q}^\prime )^2+
({\bf k}+{\bf k}^\prime )^2/4]r_0^2/4-[({\bf Q}+{\bf Q}^\prime )^2+
({\bf k}-{\bf k}^\prime )^2/4]r_1^2/4}\Bigg\}. \]
The two terms in the braces of Eq. (\ref{thirt}) correspond to the two summands in Eq. (\ref{twn}) which
describe two
types of wave correlations in the course of four waves propagation through turbulent atmosphere.
As we see, the
contributions of both trajectories are not equivalent when $r_1\not =r_0$. The effect is entirely due
to partial coherence and may be controlled by means of variation of $r_1$. For the case of $r_1=r_0$,
Eq. (\ref{thirt}) is reduced to Eq. (\ref{thiro}).

The relative contributions of the two terms in Eq. (\ref{thirt}) to
the intensity correlation function can be easily estimated. The
effective volumes of integration of the first and second terms over
${\bf q}$ and ${\bf q}^\prime$ are of the order of $r_1^{-4}$ and
$r_1^{-2}r_0^{-2}$, respectively. Moreover, for long-distance
propagation, the integrations over ${\bf k}$ and ${\bf k}^\prime $
are confined to the ``turbulent" terms, but not by $r_1^{-1}$ or
$r_0^{-1}$. Hence the second term gives a contribution, which is
$(r_0/r_1)^2$ times less than the first one. It follows from a
comparison of Eqs. (\ref{thiro}) and (\ref{thirt}) that the
intensity correlation function measured by a fast detector is
approximately twice the value of a slow-detector measurement when
$r_0>>r_1$. Of course, this estimate is only valid for long-distance
propagation. The next section deals with the case of a slow detector
in more detail.

\section{Calculations of the intensity fluctuations}

 It follows from previous considerations that the intensity correlation function for a slow
detector is given by
 \begin{equation}\label{thirth}
<:I(t+\tau )I(t):>=\Bigg(\frac {2\pi c\hbar \omega _0r_1^2}{VL_xL_y}\Bigg)^2<b^+b^+bb>\times
\end{equation}
\[ \Sigma <\Bigg\{ e^{-(Q^2+{Q^\prime }^2)r_1^2/2-(k^2+{k^\prime }^2)r_0^2/8}+
e^{-[({\bf Q}-{\bf Q}^\prime )^2+
({\bf k}+{\bf k}^\prime )^2/4]r_0^2/4-[({\bf Q}+{\bf Q}^\prime )^2+
({\bf k}-{\bf k}^\prime )^2/4]r_1^2/4}\Bigg\} \]
 \[ \times e^{-i\{{\bf k}[{\bf r}-{\bf c}({\bf q})t]+
{\bf k}^\prime [{\bf r}-{\bf c}({\bf q}^\prime )t]+(c/q_0)\int _0^tdt^\prime
t^\prime ({\bf k}{\bf F}[{\bf r}({\bf q},t^\prime )]+{\bf k}^\prime {\bf F}[{\bf r}({\bf q}^\prime ,
t^\prime )])\}}>.\]
As we see, the problem is reduced to averaging over the refractive index fluctuations and many-fold
integrations. It is worthwhile to recall here that ${\bf Q}$ and ${\bf Q}^\prime $ are dependent
on the fluctuating force ${\bf F}$. To get a linear form with respect to ${\bf F}$ in the
exponents of Eq. (\ref{thirth}), the integral representations for the factors containing ${\bf Q}$
and ${\bf Q}^\prime $ may be used:
\begin{equation}\label{thirf}
e^{-Q^2r_1^2/2}=\int \frac {d{\bf p}}{2\pi r_1^2}e^{i{\bf
pQ}-p^2/(2r_1^2)},
\end{equation}
\begin{equation}\label{thirfi}
e^{-({\bf Q}+{\bf Q}^\prime )^2r_1^2/4-({\bf Q}-{\bf Q}^\prime )^2r_0^2/4}=\int \frac {d{\bf p}
d{\bf p}^{\prime}}{(2\pi r_0r_1)^2}e^{i{\bf pQ}+i{\bf p}^\prime {\bf Q}^\prime -({\bf p}-{\bf p}
^\prime )^2/(4r_0^2)-({\bf p}+{\bf p}
^\prime )^2/(4r_1^2)}.
\end{equation}
After substitution of expressions (\ref{thirf}) and (\ref{thirfi}) into Eq. (\ref{thirth}), further
analysis
is facilitated  considerably. In this case the fluctuating field enters the intensity correlation
function only via the factor
\begin{equation}\label{thirs}
e^{-i\int_0^tdt^\prime \{ ({\bf p}+{\bf k}t^\prime s/q_0){\bf F}[{\bf r}({\bf q},t^\prime )]+
({\bf p}^\prime +{\bf k}^\prime t^\prime s/q_0){\bf F}[{\bf r}({\bf q}^\prime,t^\prime )]\} }.
\end{equation}
The averaging of Eq. (\ref{thirs}) over the refractive index fluctuations may be performed
straightforwardly if we consider
the trajectories ${\bf r}({\bf q},t^\prime )$ and ${\bf r}({\bf q}^\prime,t^\prime )$ to be
unperturbed by the turbulence, i.e. ${\bf r}({\bf q},t^\prime )={\bf r}+{\bf c}({\bf q})(t^\prime-t)$,
and ${\bf r}({\bf q}^\prime ,t^\prime )={\bf r}+{\bf c}({\bf q}^\prime )(t^\prime -t)$. Then,
the average value of Eq. (\ref{thirs}) is given by
\begin{equation}\label{thirse}
exp\Bigg\{ -0.033C_n^2\pi ^2q_0^2\int _0^zdx \int _0^\infty dgg^{-2/3}e^{-(gl_0/2\pi )^2}
[({\bf p}+{\bf k}x /q_0)^2+({\bf p}^\prime +{\bf k}^\prime x /q_0)^2
\end{equation}
\[ + 2(p_\parallel +k_\parallel x/q_0)(p_\parallel ^\prime +k_\parallel ^\prime x /q_0)
(J_0-J_2)+2(p_\perp +k_\perp x/q_0)(p_\perp ^\prime +k_\perp ^\prime x/q_0)(J_0+J_2)]\Bigg\}, \]
where $J_0$ and $J_2$ are zeroth and second order Bessel functions with the arguments
equal to $g|{\bf q}-{\bf q}^\prime |(z-x)/q_0$. The indices $\{_\parallel \}$ and $\{_\perp  \}$
indicate the parallel and perpendicular to ${\bf q}-{\bf q}^\prime $ components of the corresponding 2D
vectors. In the derivation of Eq. (\ref{thirse}) we have used the relations
\begin{equation}\label{thirei}
<{\bf F}[{\bf r}({\bf q},t^\prime )]{\bf F}[{\bf r}({\bf q},t^{\prime \prime })]>=
<{\bf F}[{\bf r}({\bf q},t^\prime )-{\bf r}({\bf q},t^{\prime \prime })]{\bf F}(0)>=
<{\bf F}[{\bf c}_{\bf q}(t^\prime -t^{\prime \prime })]{\bf F}(0)>
\end{equation}
and
\begin{equation}\label{thirni}
<{\bf F}[{\bf r}({\bf q},t^\prime )]{\bf F}[{\bf r}({\bf q}^\prime ,t^{\prime \prime })]>=
<{\bf F}[{\bf r}({\bf q},t^\prime )-{\bf r}({\bf q}^\prime ,t^{\prime \prime })]{\bf F}(0)>
\end{equation}
\[ =<{\bf F}[c_{{\bf q}^\prime }(t^\prime -t^{\prime \prime })-{\bf c}_{{\bf q}-{\bf q}^\prime }(t-t^\prime )]
{\bf F}(0)>\approx <{\bf F}[c{\bf e}_z(t^\prime -t^{\prime \prime })-{\bf c}_{{\bf q}-{\bf q}^\prime }
(t-t^\prime )]{\bf F}(0)>,\]
where ${\bf e_z}$ is a unit vector in the $z$-direction. The two first terms in square brackets of
Eq. (\ref{thirse})
describe correlations of waves with the same ${\bf q}$ and ${\bf q}^\prime $, while two other terms
describe cross-correlations. The contribution of the last terms decreases with increasing distance of
propagation, $z$, because $J_{0,2}(y)\rightarrow 0$ when $y\rightarrow \infty $. Neglecting the
cross-correlation terms, we can easily obtain the asymptotic value ($z\rightarrow \infty $) of the
intensity fluctuations and
the scintillation index. The scintillation index is given by
\begin{equation}\label{forty}
\sigma ^2=\frac {<:I(t+\tau )I(t):>-<I>^2}{<I>^2}=\frac
{r_1^2}{r_0^2}.
\end{equation}
Similar result for the limiting case $\frac {r_1}{r_0}\rightarrow 0$
was obtained in \cite{bana54}. At the same time, this result differs
from the asymptotic value equal to 1, obtained in \cite{korot04}.
The difference may arise because of a different assumptions on the
coherent properties of the sourse used by the authors of
\cite{korot04}. Namely, in \cite{korot04} the authors consider a
small coherence time equal to inverse bandwidth of the laser
generation, while our results are valid for small characteristic
times of local phase fluctuations introduced by the dynamic phase
screen.

As we see, the scintillation index tends to zero when $z\rightarrow \infty ,(r_1/r_0)^2\rightarrow 0$.
This property of partially coherent radiation is favorable for practical utilization.

Similar reasonings may be used to obtain the correlator of arbitrary
(the $m$th) order. It is given by
\begin{equation}\label{fortyo}
\frac {<:\prod _{i=1}^{m}[I(t_i)-<I>]:>}{<I>^{m}} =a_m\Bigg(\frac {r_1}{r_0}\Bigg)^{m},
\end{equation}
where $\tau_s<|t_i-t_j|<\tau_d$ and $a_m=1,2,9,44,265..$ for
$m=2,3,4,5,6..$, respectively.

Taking into account Eqs. (\ref{thirth})-(\ref{thirs}) and all terms
in Eq. (\ref{thirse}) we obtain an analytic expression for the
intensity fluctuations in which many integrations can be performed
analytically. The rest (three-fold integral) can be evaluated
numerically. This allows us to analyze the effect of turbulence on
$\sigma ^2$ at large but finite distances.

Eq. (\ref{thirse}) is derived under the assumption that the
trajectories ${\bf r}({\bf q},t^\prime)$ in the fluctuating force
$F$ are straight lines. Let us analyse the effect of distortion of
trajectories due to the fluctuating force. With regard for the fact
that the force $\bf F$ is responsible for changing the photon
transverse momentum (and the transverse velocity), the account for
the dependence of $\bf F$ on $\bf F$ means the account of nonlinear
in $\bf F$ effects in the photon transverse displacements. These
effects are beyond the applicability of the fourth-moment equation
\cite{tatar},  \cite{fant75}, therefore, their studies are
impossible by means of the Yakushkin method. Also, this means that
papers \cite{bana54}, \cite{bana55}, which are based on the
Yakushkin method, stay this problem out of consideration.

First of all, it should be noted that the correlation function
$<{\bf F}({\bf r}_1){\bf F}({\bf r}_2)>$ depends on ${\bf r}_1-{\bf r}_2$. If both points belong
to the same trajectory
[${\bf r}_1\equiv  {\bf r}({\bf q},t^\prime ),{\bf r}_2\equiv  {\bf r}({\bf q},t^{\prime \prime })$]
this difference is equal to
${\bf r}({\bf q},t^\prime )-{\bf r}({\bf q},t^{\prime \prime })
\approx c{\bf e}_z(t^\prime -t^{\prime \prime})$. [See Eq. (\ref{thirei}).] Consequently
$|t^\prime -t^{\prime \prime }|\leq L_0/c$  and only negligible particle displacements in
perpendicular to the $z$-direction may occur for such very short time intervals. In the other case, when
two points belong to different trajectories [${\bf r}_1\equiv  {\bf r}({\bf q},t^\prime ),
{\bf r}_2\equiv {\bf r}({\bf q}^\prime ,t^{\prime \prime })$] we have
\begin{equation}\label{fortyt}
{\bf r}_1-{\bf r}_2\approx c{\bf e}_z(t^\prime -t^{\prime \prime})-{\bf c}_{{\bf q}-{\bf q}^\prime}
(t-t^\prime )+\frac c{q_0}\int _{t^\prime }^tdt_1(t_1-t^\prime )\{ {\bf F}[{\bf r}({\bf q},t_1)]-
{\bf F}[{\bf r}({\bf q}^\prime ,t_1)]\} .
\end{equation}
It can be easily seen from Eq. (\ref{fortyt}) that for any given
$|t^\prime-t^{\prime \prime}|\leq L_0/c$, the last two terms on the
right-hand side may be comparable with the first term for
sufficiently large values of $t-t^\prime $. In this case, the
deviations of the trajectories from straight lines, which enter
arguments in the left hand side of Eq. (\ref{thirni}), should be
taken into account. One can say that the deviations are accumulated
throughout the whole propagation path and, in contrast to the case
of the same trajectory, may become sufficiently large to influence
the wave correlations.

For further analysis, an important property of the correlation
function $<{\bf F}({\bf r}_1-{\bf r}_2){\bf F}>$, where ${\bf
r}_1-{\bf r}_2$ is given by Eq. (\ref{fortyt}), has to be noted.
There is almost no correlation between ${\bf F}$
 entering the argument of ${\bf F}$ and the function itself. This is because of the negligibly small
time intervals (or characteristic distances) where these functions correlate:
$0<t_1-t^\prime, t^{\prime \prime}\leq L_0/c$. Then, if we consider as previously that $L_0<<z$, the
averaging of $<{\bf F}({\bf r}_1-{\bf r}_2){\bf F}>$ may be undertaken in two steps: firstly,
we average this quantity considering ${\bf r}_1-{\bf r}_2$ as a fixed parameter and after that the
remaining averaging should be performed. The result is
 \begin{equation}\label{fortyth}
<F_i({\bf r}_1-{\bf r}_2)F_j(0)>=\int d{\bf g}g_ig_j\psi ({\bf g})<e^{-i{\bf g}({\bf r}_1-{\bf r}_2)}> ,
\end{equation}
where ${\bf g}$ is a three-dimensional vector and the indices $i,j$
denote the components perpendicular to $z$. The procedure described
above may be repeated to obtain $<e^{-i{\bf g}({\bf r}_1-{\bf
r}_2)}>$. Again, this function may be expressed in terms of
two-point correlation functions in which the points belong to the
same or different trajectories. At this stage we will simplify
further analysis by imposing the approximation
\begin{equation}\label{fortyfo}
<e^{-i{\bf g}({\bf r}_1-{\bf r}_2)}>\approx <e^{-i{\bf g}{\bf r}_1}><e^{i{\bf g}{\bf r}_2}> ,
\end{equation}
where the correlation of different trajectories is neglected.
Physically, this approximation requires that two  photons with
different transverse momenta propagate in spatial areas with
different refractive indices and experience different fluctuating
forces. Therefore they gain different values of the transverse
velocity and transverse displacements. It is evident that for any
finite value of ${\bf q}-{\bf q}^\prime $, the displacements become
uncorrelated when $z\rightarrow \infty $.

Within the approximation (\ref{fortyfo}), the effect of refractive index fluctuations on trajectories
in Eq. (\ref{thirs}) may be accounted for by multiplying all the Bessel functions in Eq. (\ref{thirse})
by the factor $e^{-\gamma (z-x)^3}$, where
\[ \gamma =\gamma (g)=0.011\pi ^2(2\pi )^{1/3}\Gamma \Bigg(\frac 16\Bigg)C_n^2l_0^{-1/3}g^2.\]
This factor reduces the effect of correlations of different trajectories by taking into account
randomization of the particle displacements from straight lines.

\section{Weak irradiance fluctuations}

For the case of weak turbulence or short distance propagation, the beam characteristics are the same
as for propagation in free space. Small deviations from free space regime may be
accounted using perturbation methods. As previously, the intensity fluctuations may be described in
terms of the function $G({\bf q},{\bf q}^\prime ;{\bf k},{\bf k}^\prime )$. Further analysis is
simplified
if one uses an iterative procedure not for $G$, but for its fluctuating part, $\Upsilon $,
defined as
\begin{equation}\label{fortyfi}
\Upsilon ({\bf q},{\bf q}^\prime ;{\bf k},{\bf k}^\prime )=G({\bf q},{\bf q}^\prime ;{\bf k}{\bf k}^
\prime )-<b^+_{{\bf q}+{\bf k}/2}b_{{\bf q}-{\bf k}/2}><b^+_{{\bf q}^\prime +{\bf k}^\prime /2}
b_{{\bf q}^\prime -{\bf k}^\prime /2}>.
\end{equation}
The equation of motion for $\Upsilon $ is given by
\begin{equation}\label{fortys}
[\partial _t-i\Delta ({\bf q},{\bf q}^\prime ;{\bf k},{\bf k}^\prime )]
\Upsilon ({\bf q},{\bf q}^\prime ;{\bf k},{\bf k}^\prime )=i\omega _0\sum _{\bf g}n_{\bf g}\{
\Upsilon ({\bf q}+{\bf g}/2,{\bf q}^\prime ;{\bf k}-{\bf g},{\bf k}^\prime )
\end{equation}
\[ -\Upsilon ({\bf q}-{\bf g}/2,{\bf q}^\prime ;{\bf k}-{\bf g},{\bf k}^\prime) +[{\bf q},{\bf k}
\leftrightarrow {\bf q}^\prime ,{\bf k}^\prime ]\} , \]
where the symbol $[{\bf q},{\bf k}\leftrightarrow {\bf q}^\prime ,{\bf k}^\prime ]$ indicates
two summands,
which are similar to previous ones but with the variables interchanged as indicated, and
\[ \Delta ({\bf q},{\bf q}^\prime ;{\bf k},{\bf k}^\prime )\equiv \omega _{{\bf q}+{\bf k}/2}+
\omega _{{\bf q}^\prime +{\bf k}^\prime /2}-\omega _{{\bf q}^\prime -{\bf k}^\prime /2}-
\omega _{{\bf q}-{\bf k}/2}.\]  It follows from
Eq. (\ref {fortys}) that the average value of $\Upsilon $ may be written as
\begin{equation}\label{fortyse}
<\Upsilon ({\bf q},{\bf q}^\prime ;{\bf k},{\bf k}^\prime )>=i\omega _0\sum _{\bf g}\int
_0^tdt^
\prime e^{i\Delta ({\bf q},{\bf q}^\prime ;{\bf k},{\bf k}^\prime )(t-t^\prime )}
\end{equation}
\[ \times <n_{\bf g} \{ \Upsilon ({\bf q}+{\bf g}/2,{\bf q}^\prime ;{\bf k}-{\bf g},{\bf k}^\prime)-
\Upsilon ({\bf q}-{\bf g}/2,{\bf q}^\prime ;{\bf k}-{\bf g},{\bf k}^\prime )
+[{\bf q},{\bf k}\leftrightarrow {\bf q}^\prime ,{\bf k}^\prime ]\}|_{t=t^\prime}>, \]
where the initial condition $<\Upsilon ({\bf q},{\bf q}^\prime ;{\bf k},{\bf k}^\prime )>|_{t=0}=0$
corresponding to the case of slow detector, was used.
The right-hand side of Eq. (\ref{fortyse}), dependent on average value of $<n_{\bf g}\Upsilon >$, is
still unknown. An expression similar to Eq. (\ref{fortyse}) may be derived again for
 $<n_{\bf g}\Upsilon >$.
Then, we will simplify the problem by considering the time dependence of all operators in the
integrand to be
determined by free-space propagation laws. The approach is in the spirit of the classical Rytov
approximation \cite{tatar}. This will make it possible to integrate all terms and to get
explicit forms for $<n_{\bf g}\Upsilon >$ as well as for $<\Upsilon >$. Using some simple algebra,
it can be proved that the scintillation index is given by
\begin{equation}\label{fortyei}
\sigma ^2=\sigma _1^2K(z,r_0,r_1),
\end{equation}
where $\sigma _1^2$ is the Rytov variance defined by $\sigma _1^2=1.23C_n^2q_0^{7/6}z^{11/6}$ and
\begin{equation}\label{fortyn}
K(z,\rho_0,\rho_1)=4.24\int _0^1d\tau \int _0^\infty dxx^{-8/3}exp\Bigg\{-x^2\Bigg[\frac
{q_0l_0^2}{4\pi ^2z}+
\tau ^2\frac {\rho_0^2+\rho_1^2}{4+\rho_0^2\rho_1^2}\Bigg]\Bigg\}
\end{equation}
\[\times sin^2\Bigg(\frac {\tau x^2}2-\frac {2\tau ^2x^2}{4+\rho _0^2\rho _1^2}\Bigg),\]
where
$\rho _{0,1}^2= {r_{0,1}^2q_0}/z $. As done previously, in course of derivation of Eq. (\ref{fortyei}),
we have considered the propagation distance to be much longer than the characteristic scale of
turbulence variation ($z\gg L_0$).
\begin{figure}[ht]
\centering
\includegraphics{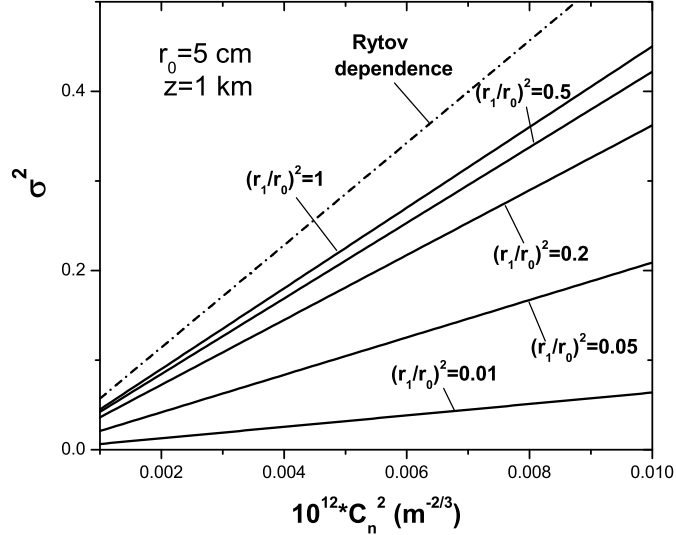}
\caption{Dependence of scintillation index on turbulence strength for weak fluctuations; $q_0$ and
$l_0$ are as in Fig. 1.
Decreasing of scintillation index with the decrease of source coherence is seen (similar to
results of O.~Korotkova {\it et al.} \cite{korot04}). }
\label{here_is_the_label}
\end{figure}
It follows from Eq. (\ref{fortyn}) that in the limit of $\rho
_1\rightarrow \infty , l_0\rightarrow 0$, we have the result of
Rytov theory ($\sigma ^2=\sigma _1^2$) because $K\rightarrow 1$. The
quantity in the square brackets of Eq. (\ref{fortyn}) is negligible
in this case. With decreasing initial beam coherence, $r_1$ becomes
smaller and it may occur that the quantity in square brackets
becomes sufficiently large to influence the result of integration.
The character of wave propagation becomes modified from the plane
wave regime to the spherical wave regime. This is accompanied by a
decrease of $\sigma ^2$. The effect is saturated at some small
values of $r_1\sim 2z/r_0q_0$ and further decrease of the coherent
length $\lambda _c$ has no effect on $\sigma ^2$. Also, when $\rho
_0$ is small, the effect of $r_1$ variation is of no significance.
Therefore, there is the opportunity to control $\sigma ^2$ but this
is only possible at a sufficiently large aperture radius $r_0$ and
small distance $z$. Furthermore, the reduction of $\sigma ^2$ is
limited by some finite value. Fig. 2 illustrates the effect of
partial coherence just for the favorable case when $\rho_0=5$.

\section{Discussion}

Figs. 3 and 4 show the dependence of scintillation index on the turbulence for two different distances,
$z$. A well-pronounced effect for the
decreasing of $\sigma ^2$ with the decrease of the initial coherence can be seen in the range of strong
turbulence.
\begin{figure}[ht]
\centering
\includegraphics{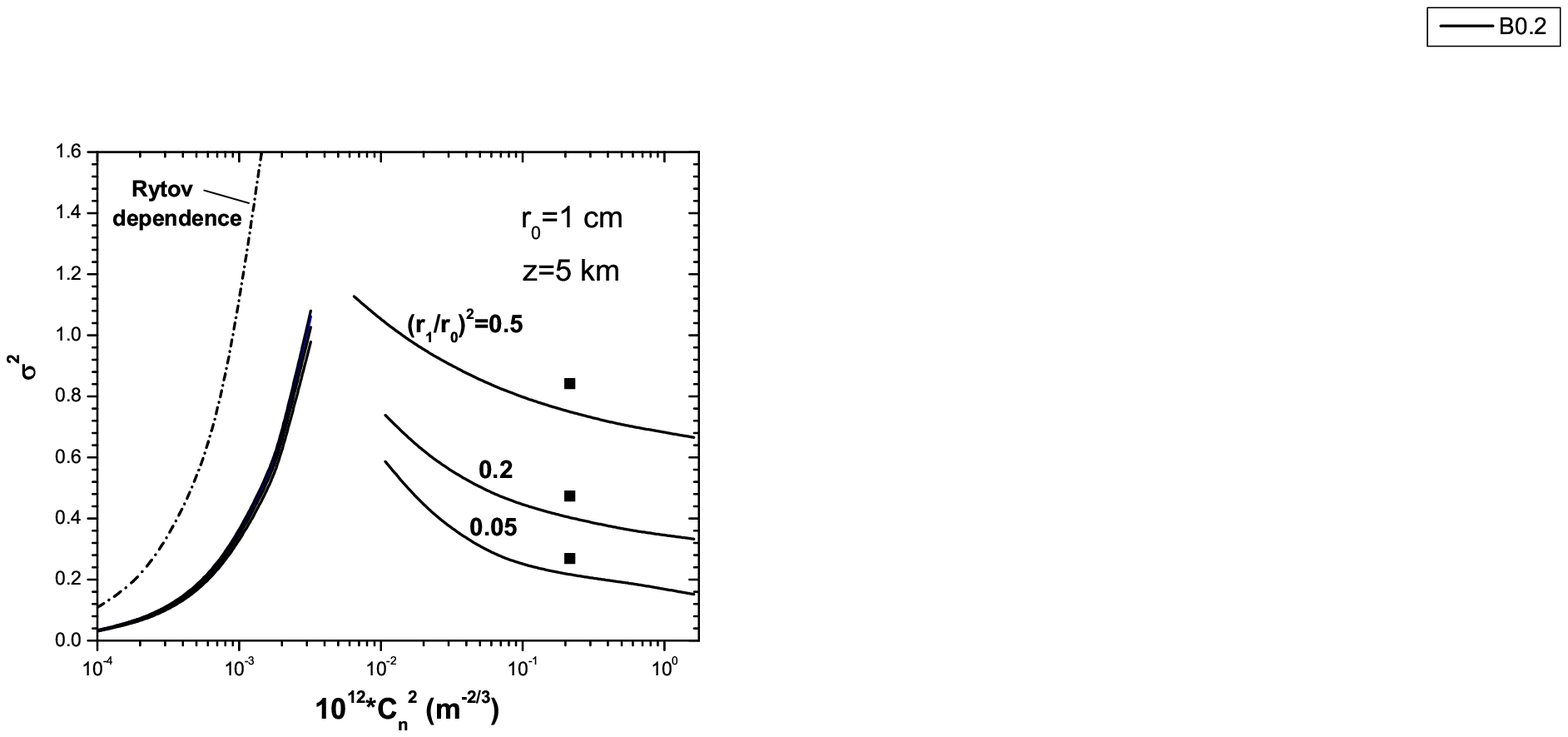}
\caption{Plots $\sigma ^2(C_n^2)$ for different initial coherence $(r_1/r_0)^2$; $q_0$ and
$l_0$ are as in Fig. 1.
Left-side curves calculated with employing Eqs. (\ref{fortyei}) and (\ref{fortyn}) exhibit negligible
effect of partial coherence (merged curves), that is in contrast to
results shown in Fig. 2. Two very different parameters $\rho_0^2 $ are used in these cases. Black squares
show $\sigma ^2$ obtained with the assumption of straight trajectories in Eq. (\ref{thirs}).}
\label{here_is_the_label}
\end{figure}
 There is a very simple physical explanation of the reduction of $\sigma ^2$. Two things are
very important for understanding
this phenomenon. First of all, in the course of the irradiance propagation the beam acquires the
properties of Gaussian statistics. Therefore, the asymptotic value of the intensity correlations
 $<I(t)I(t)>$ is given by $2<I>^2$. On the other hand, the quadratic detector counts are determined
not by simultaneous correlations, but by the average $<I(t)I(t+\tau
)>$, where the characteristic value of $\tau $ is of the order of
$\tau_d$. In the limiting case of $\lambda_c\rightarrow 0$ and $\tau
>>\tau_s$ (slow detector), there are no correlations between $I(t)$
and $I(t+\tau )$. Hence, $<I(t)I(t+\tau )>=<~I>^2$, and the
normalized variance of the intensity fluctuations is negligible
($\sigma ^2\rightarrow 0$). This physical picture is quite similar
to the well-known Hanbary-Braun-Twiss effect \cite{mand}, or photon
bunching for thermal light: at zero delay the correlation function
has twice the value for long delays.
\begin{figure}[ht]
\centering
\includegraphics{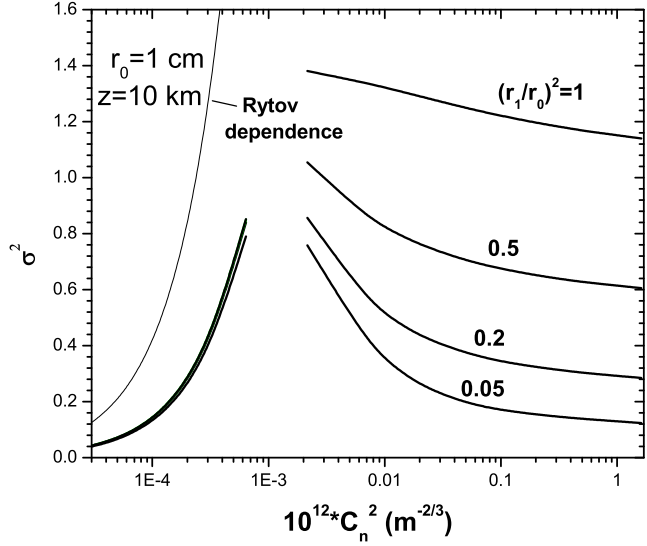}
\caption{The same as in Fig. 3 but for longer distance. The upper curve corresponds to coherent
beam. It approaches asymptotically to value $\sigma ^2=1$.}
\end{figure}
For finite values of $\lambda _c$, the irradiance differs from
thermal light resulting in a finite value of the scintillation
index. Our theory gives the scintillation index equal to
$r_1^2/r_0^2$, that is the relative part of the aperture where the
exiting light may be considered as a coherent light.

It should be noted that the finite value of $\sigma ^2$ shows an
absence of full thermalization because the memory of the photon flux
about source correlations still exists. This memory may be lost due
to fluctuations of the photon transit time $\delta t$. [The transit
time was assumed to be constant
 (equal to $z/c$) in the previous analysis.] When $\delta t \not =0$,
 the initial conditions for $f({\bf r},{\bf q},t)$
correspond not to its value at the aperture plane, but, for example,
to some outlying point $\delta z =c|\delta t|$ where the turbulence
has already modified the waves. The estimate of $\delta z$ is given
by its rms value. Setting
\[ \delta z=\int _0^zdz^\prime \delta n(z^\prime),\]
we can easily obtain $<(\delta z)^2>\sim 0.066\pi
^2zC_n^2L_0^{5/3}$. When $C_n^2=10^{-13}m^{-2/3}$,  $L_0=10m,
z=10^5m $, we have $<(\delta z)^2>^{1/2}\sim 0.55*10^{-3}m$ and
$|\delta t|\sim 2*10^{-12}s$, which is negligibly small.

Other causes of transit time fluctuations arise from fluctuations of the
transverse velocity of the photons. For nonzero $q_\bot $ the photon
velocity in the $z$ direction is given by
\begin{equation}\label{fifty}
 c_z=\frac {\partial \omega _{\bf q}}{\partial q_z}\approx c\Bigg(1-\frac {q_\bot ^2}{2q_0^2}\Bigg).
\end{equation}
Assuming $q_\bot $ to be caused by the turbulence, we have
 \begin{equation}\label{fiftyo}
{\bf q}(z) = \int _0^z{\bf F}(z^\prime )dz^\prime /c,
\end{equation}
where indices $_\bot$ are omitted again. Then, using Eqs. (\ref{fifty}) and
(\ref{fiftyo}) we may estimate $\delta t$ as
\begin{equation}\label{fiftyt}
 \delta t\sim \int_0^z\frac {dz^\prime }c\Bigg<\frac 1{1-q^2/2q_0^2}-1\Bigg>\approx \int _0^z\frac
{dz^\prime }{c}\frac {<q^2(z^\prime )>}{2q_0^2}\approx 1.7\frac {z^2C_n^2}{cl_0^{1/3}}.
\end{equation}
For the same parameters used previously, $\delta t \sim 3*10^{-11}s$ and $<(\delta z)^2>^{1/2}\approx
9*10^{-3}m$, which is negligible again.

The most serious assumption of our approach is that the influence of
the turbulence on the photon distribution can be described in terms
of a random force ${\bf F}({\bf r})$ (see Eq. (\ref{seven}))
modifying photon momentum ${\bf q}\bot z$. Mathematically, this
approximation can be justified when the width of the photon
distribution in momentum space is greater than those turbulence wave
vectors, which give the dominant contribution to $\sigma ^2$. In the
vicinity of the source, the photon momentum  is distributed within
the range of the order of $\pi /r_1$. (See Eq. (\ref{twth}).) The
turbulence spectrum covers very broad interval of wave vectors $[\pi
/L_0,\pi /l_0]$ where $L_0/l_0\sim 10^5$. Therefore, it is not clear
a priori what wave vector should be taken for the comparison. Of
course, the condition $r_1\leq l_0$ is sufficient to validate our
approach. At the same time, this condition imposes very rigid
restriction on real optical systems. Fortunately, it is not
obligatory for a reliable solution for long-distance propagation.
Fist of all, it may be assumed that a broad range of the turbulence
spectrum, rather than wave vectors equal to $\pi/l_0$, contributes
significantly to measured quantities. If this is true, these
quantities would not be sensitive to the boundary values of wave
vectors. In this context, it is important to note that the beam
radius really does exhibit a weak dependence on $l_0$. [It depends
on $l_0$ through $T\sim l_0^{-1/3}$, see Eq. (\ref{tws}).] Hence, it
is reasonable  to consider the characteristic value of the
turbulence wave vector to be much smaller than $\pi /l_0$.

On the other hand, due to the action of the random force, there is a diffusion-like
increase in the transverse momentum of photons in the course of their propagation. It is
just that quantity (not $\pi/r_1$) which should be compared with the turbulence wave vectors for
large $z$. Using Eq. (\ref{fiftyo}) we may estimate the increase of the transverse momentum as
 \begin{equation}\label{fiftyth}
 <q^2>=0.066{\pi }^2\Gamma \Bigg(\frac 16\Bigg)q_0^2zC_n^2(2\pi /l_0)^{1/3}.
\end{equation}
Substituting the previous parameters into Eq. (\ref{fiftyth}), we get $<q^2>/{\pi }^2l_0^{-2}\approx 10^2$.
This estimate shows our approach to be reliable for long distance propagation. At small distances,
the perturbation theory is applicable.

\section{Conclusion}
The approach presented in this paper can be used for both stationary
beams as well as for beams with varying intensity. Also, the
statistics of the exiting irradiance may differ from the statistics
of coherent or partially coherent beams. Our method of obtaining the
correlation function for the intensity fluctuations is not based on
the solution of an equation for the fourth-order correlation
function. Moreover, we have obtained the asymptotic value for all
moments of the intensity fluctuations. Our method does not employ
Markov approximation used previously to derive the equation for the
fourth-order correlation function. Also, we do not use the so-called
quadratic approximation, which like the Markov approximation is
nothing more than an artificial modification of the refractive index
covariance in order to simplify the theoretical treatment.  Both
simplifications are inherent to papers \cite{bana54} and
\cite{bana55}. Besides that, there was assumed $l_0=0$, that is
impossible within our consideration. Suffice it to say that in the
case of $l_0=0$, the quantities $T$ and $\gamma $ lose their
physical sense because of infinitely large values. In spite of the
presence of so serious approximations, the asymptotic ($z\rightarrow
\infty$ or $C_n^2 \rightarrow \infty $) results of \cite{bana54} and
our paper results are very similar. The explanation of this follows
from the result obtained: the asymptotic value of $\sigma ^2$ does
not depend on the atmosphere turbulence mechanism. Therefore, it
does not matter what kind of turbulence spectrum is employed in the
case of saturated fluctuations. At the same time, our calculations
show (see Figs. 3 and 4) that the distinct asymptotic regime
scarcely can be reached in real experiments. Thus, for practice, it
is more important to know the behavior of $\sigma ^2$ at large but
finite $C_n^2$ and $z$ than its asymptotic value. In this case, the
real turbulence spectrum as well as an adequate theoretical analysis
are required. In this context, the motivations of our studies are
evident.

Our calculations show how to suppress the intensity fluctuations
using a PCB. The conditions required for this are a random phase
modulation of the signal and the use of a slow detector. Estimations
of the utility of the PCB should take into account the negative
factor of worsening for the receiving system resolution when $\tau
_d$ is increased. Also, it is important for the experiment
performance, that the ratio $\tau_s/\tau_d$ is always finite, and
just this ratio rather than $(r_1/r_0)^2$ determines the asymptotic
value of $\sigma $ when $(\tau_s/\tau_d)>(r_1/r_0)^2$. Hence, the
regime of fast phase modulations (but not the increase of $\tau_d$)
is preferable in both cases.

\section{Acknowledgment}
We are grateful to V.N. Gorshkov  for helping us with illustrative
material, and to L.C. Andrews and B.M. Chernobrod for useful
discussions. This work was supported by the Department of Energy
(DOE) under Contract No. W-7405-ENG-36.

\newpage \parindent 0 cm \parskip=5mm





\end{document}